\documentclass[conference]{IEEEtran}
\ifCLASSINFOpdf
\else
\fi
\usepackage{amsmath,dsfont,bbm,epsfig,amssymb,amsfonts,amstext,verbatim,amsopn,cite,subfigure,multirow,multicol,lipsum,xfrac}
\usepackage{amsthm}
\usepackage{mathtools,amsthm}
\usepackage{perpage}
\usepackage{balance}
\usepackage{url}
\usepackage{amsfonts}
\usepackage{epsfig}
\usepackage[font={small}]{caption}
\usepackage{etoolbox}
\usepackage{algorithmicx}
\usepackage[Algorithm,ruled]{algorithm}
\usepackage{algpseudocode}
\usepackage{pifont}
\usepackage[utf8]{inputenc}
\usepackage[T1]{fontenc}  
\usepackage[nolist]{acronym}
\MakePerPage{footnote}
\usepackage{paralist}
\usepackage{enumitem}
\usepackage{bbm}
\usepackage{bm}
\usepackage[process=auto]{pstool}
\usepackage{tikz,pgfplots}
\newcommand{\snr}{E_{\rm b}/N_0}

\DeclareMathOperator*{\argmax}{arg\,max}

\newcommand{\bh}{{\mathbf{h}}}

\newcommand{\set}[1]{\left\lbrace#1\right\rbrace}

\newcommand{\mH}{\mathbf{H}}

\newcommand{\norm}[1]{\lVert #1 \rVert}

\newtheoremstyle{mystyle}
  {}
  {}
  {\it}
  {}
  {\bfseries}
  {:}
  { }
  {}

\theoremstyle{mystyle}

\newtheorem{tas}{Algorithm}
\newcounter{bar}

\begin{document}
\title{A Fair Comparison Between Spatial Modulation and Antenna Selection in Massive MIMO Systems\vspace*{4mm}}

\author{
\IEEEauthorblockN{
Bernhard G{\"a}de,
Ali Bereyhi,
Saba Asaad,
Ralf R. M\"uller,
}
\IEEEauthorblockA{
Institute for Digital Communications (IDC), Friedrich-Alexander-Universit\"at Erlangen-N\"urnberg (FAU)\\
\{bernhard.gaede, ali.bereyhi, saba.asaad, ralf.r.mueller\}@fau.de\vspace*{-2mm}
\thanks{
This work has been accepted for presentation in the 23rd International ITG Workshop on Smart Antennas (WSA 2019) in Vienna, Austria. The link~to~the final version in the Proceedings of WSA will be available later.
}
}
}


\IEEEoverridecommandlockouts

\maketitle

\begin{abstract}
Both antenna selection and spatial modulation allow for low-complexity MIMO transmitters when the number of RF chains is much lower than the number of transmit antennas. In this manuscript, we present a quantitative~performance~comparison between these two approaches by taking into account implementational restrictions, such as antenna switching. 

We consider a band-limited MIMO system, for which the pulse shape is designed, such that the outband emission satisfies~a~desired spectral mask. The bit error rate is determined for~this~system, considering antenna selection and spatial modulation. The results depict that for any array size at the transmit and receive sides, antenna selection outperforms spatial modulation, as long as the power efficiency is smaller than a certain threshold level. By passing this threshold, spatial modulation starts to perform superior. Our investigations show that the threshold takes smaller values, as the number of receive antennas grows large.~This~indi-cates that spatial modulation is an effective technique for uplink transmission in massive MIMO systems. %
\end{abstract}


%

\section{Introduction}
Recently, \ac{sm} has been proposed for single \ac{rf} chain multi-antenna transmission. This technique addresses various  limitations~in \ac{mimo} systems without significant degradation of the performance \cite{basar2016index}. The main drawback of conventional \ac{mimo} techniques arises from the non-negligible cost-complexity issue which is due to
\begin{inparaenum}
\item inter-channel interference caused by multiple spatial symbols, 
\item requiring strict~synchronization among transmitting antennas, and
\item employment of a dedicated \ac{rf} chain per transmit antenna \cite{mesleh2008spatial, jeganathan2009space}.
\end{inparaenum}

\ac{sm} is a novel digital modulation technique in which the \textit{indexes} of active transmit antennas are	 used as means to convey additional information leading to higher spectral efficiency.
More precisely, the incoming bit stream is mapped into two sub-blocks called the \textit{signal constellation diagram} and \textit{spatial constellation diagram}. The first sub-block is used to select a symbol from the signal constellation, e.g., phase shift keying, and the second sub-block specifies the position of the active transmit antenna. In the basic form of  \ac{sm}, only  one of the transmit antenna is permitted to be active in each channel use. In addition to the overall complexity reduction, this approach leads to elimination of inter-channel interference. As a result, synchronisation of the transmit antennas is not required in this case.

\subsection*{Antenna selection vs. Spatial Modulation}
An alternative approach to mitigate the hardware complexity in \ac{mimo} settings is \ac{as} \cite{molisch2003mimo,molisch2005capacity}. Due to high implementational cost and complexity of massive \ac{mimo} setups \cite{larsson2014massive,hoydis2013massive}, \ac{as} has received a great deal of attention in the context of massive \ac{mimo} systems; see for example \cite{bereyhi2018stepwise,asaad2018massive,li2014energy} and the references therein.

In contrast to \ac{sm}, \ac{as} selects the transmit antennas based on the \ac{csi}. Hence, with a given constellation, the number of information bits transmitted in each channel use via \ac{as} is less than the one achieved by \ac{sm}. Despite this degradation, \ac{as} enjoys several advantages compared to \ac{sm}, such as
\begin{itemize}
\item Since the selected antenna is chosen based on \ac{csi}, \ac{as} provides diversity at transmit side.
\item In contrast to \ac{sm}, \ac{as} does not require antenna switching at each transmission interval. Therefore, the impairments caused by switching are less significant under \ac{as}.
\end{itemize}

The above discussions bring this question into mind: Given a massive \ac{mimo} setting with a single \ac{rf} chain at the transmitter, which approach performs superior, when the time-frequency resources are strictly constrained? In this manuscript, we try to answer this question. Our investigations demonstrate that for any number of transmit and receive antennas, there exists a certain level of power efficiency, before which \ac{as} outperforms \ac{sm}. As the energy efficiency passes this threshold, \ac{sm} starts to perform superior. It is further shown that by increasing the size of the receive antenna array, the threshold power efficiency becomes smaller. This observation indicates that \ac{sm} is an effective technique for uplink transmission in massive \ac{mimo} settings.

\section{System Model}
\label{sec:SysMod}
A \ac{mimo} setting with $N_{\rm t}$ transmit and $N_{\rm r}$ receive antennas is considered. The transmitter is equipped with a single \ac{rf} chain and a switching network which connects the \ac{rf} chain to a desired transmit antenna at each transmission time interval. This means that at each time interval, only one of the transmit antennas is active. Hence, the receive signal at time interval $k$ is given by
\begin{align}
\bm{y}[k] = \left.\left. \mathbf{H} \right. \bm{b}[k] \right. x[k] + \bm{w}[k].
\end{align}
\clearpage
\noindent
In the above equation, 
\begin{itemize}
\item $x[k]$ denotes the transmit symbol at time interval $k$ drawn from a modulation alphabet $\mathcal{A}$. 
\item $\bm{b}[k] \in \lbrace1,0\rbrace^{N_{\rm t}}$ describes the switching network in~time interval $k$. We refer to this vector as the \textit{selection vector}. It comprises a non-zero entry which corresponds to the active transmit antenna in interval $k$. The selection vector is in general time varying, as the switching network is allowed to switch from one antenna to another, at the beginning of each transmission time interval.
\item $\mathbf{H} \in \mathbb{C}^{N_{\rm r}\times N_{\rm t}}$ represents the matrix of channel gains. The channel experiences frequency-flat Rayleigh fading, and has a coherence interval which comprises $K_{\rm C}$ transmission intervals. Hence, the entries of $\mH$ are assumed to be \ac{iid} zero-mean complex Gaussian random variables with unit variance. 
\item $\bm{w}[k]$ is additive white complex Gaussian noise with zero mean and variance $\sigma^2$.
\end{itemize}
We assume that the transmission is performed in \ac{tdd} mode, and hence the channel is reciprocal. The \ac{csi} is further assumed to be perfectly known at both ends.

\begin{figure*}[t]
\centering
\includegraphics[width=.65\textwidth]{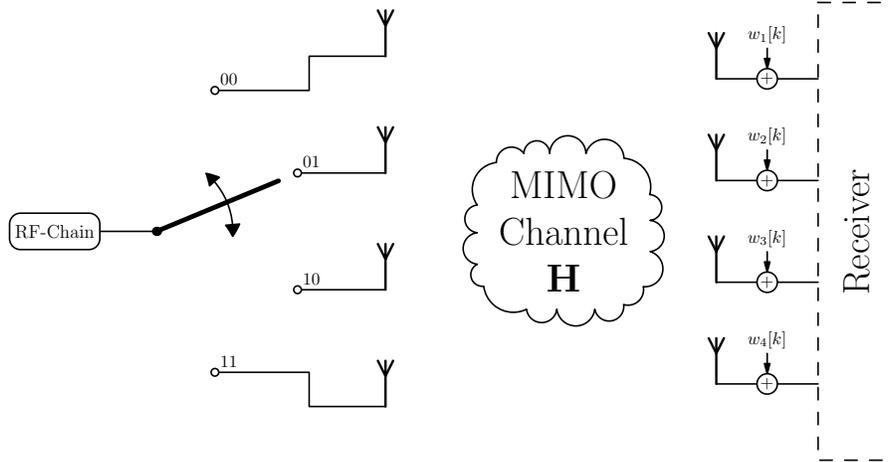}
\caption{An example of index modulation via a single \ac{rf} chain and four transmit antennas.}
\label{fig::sysMod} 
\end{figure*}

\subsection{Transmission via AS}
Under \ac{as}, the transmit antenna is chosen via a selection algorithm which optimizes a desired performance metric, e.g. the average \ac{snr} at the receive side, the achievable rate, or error rate. In the sequel, we consider a conventional selection algorithm. To illustrate the algorithm, let us write the channel matrix as
\begin{align}
\mH = \left[ \bh_1, \ldots, \bh_{N_{\rm t}} \right]
\end{align}
where $\bh_n \in \mathbb{C}^{N_{\rm r}}$ for $n\in \set{1,\ldots,N_{\rm t} }$ denotes the vector of channel gains between $n$-th transmit antenna and the receive antenna array. The \ac{as} at the transmit side is performed via the following algorithm:
\begin{tas}
\label{alg:1}
The transmitter selects the antenna whose corresponding channel gain is maximum. This means that transmit antenna $n^\star\in \set{1,\ldots,N_{\rm t} }$ is selected for which we have
\begin{align*}
n^\star = \argmax_{n\in \set{1,\ldots,N_{\rm t} }} \norm{\bh_n}^2.
\end{align*}
\end{tas}
It is worth to note that the \ac{snr} at the receive side is linearly proportional to the channel gain of the transmit antenna.~This means that the antenna selected by Algorithm~\ref{alg:1} maximizes the receive \ac{snr}. Since the selection is only based on \ac{csi}, which is known at the receive side, the receiver can infer $n^\star$, as well.

The selection vector is constructed in this case by setting entry $n^\star$ to $1$ and all other entries zero. Noting that $\bm{w}[k]$ is a stationary process and $\mathbf{H}$ is almost constant over a coherence interval, $\bm{b}[k]$ is updated only once per coherence interval when the transmitter employs \ac{as}. To reflect this, the dependence of $\bm{b} [k]$ on the transmission interval, i.e. argument $k$, is dropped in the sequel under \ac{as}.

At the receive side the \ac{ml} algorithm is utilized for detection of the transmit symbol sequence. This means that the receiver recovers symbol $x[k]$ as
\begin{align} 
\hat{x}[k] = \underset{v \in \mathcal{A}}{\argmax}\left\lbrace\left|\bm{y}[k] - \mathbf{H}\bm{b} v\right|^2\right\rbrace.
\end{align}
Noting that $\bm{b}$ has a single non-zero entry and is known at the receive side, the \ac{ml} detection algorithm under \ac{as} reduces to maximal-ratio combining.

\subsection{Transmission via SM}
\label{sec::spatialModulation}
\ac{sm} makes use of the fact that the choice of a particular antenna out of the set of available transmit antennas can also convey information when the \textit{index} of the transmit antenna is selected via the information bits. To illustrate the idea further, let us assume that the number of transmit antennas is a power of two. In this case, the indices of the transmit antennas, i.e., $n\in\set{1,\ldots,N_{\rm t}}$, is represented via $\log_2 N_{\rm t}$ bits. As a result, at each transmission interval, extra $\log_2 N_{\rm t}$ bits of information, in addition to $x[k]$, are transmitted when we select the index of the active transmit antenna via the data sequence. Fig.~\ref{fig::sysMod} shows an example of \ac{sm} with four transmit antennas. Here, the first transmit antenna is selected in transmission interval $k$, when the first two bits of data sequence, transmitted in this interval, are $0 0$. In this particular example, extra information of $\log_2 4 = 2$ bits is transmitted in each interval by \textit{index modulation}, compared to the case with \ac{as}.

In \ac{sm}, the selection vector is chosen by the data bits, which vary from one transmission interval to another. This means that, in contrast to \ac{as}, $\bm{b}[k]$ in this case varies with respect to $k$, and is not known at the receive side. %
The \ac{ml} detection, hence, requires to recover both $\bm{b}[k]$ and $x[k]$. As a result, the detection algorithm in this case reads
\begin{align}
\hat{x}[k], \hat{\bm{b}}[k] = \underset{\begin{subarray}{c}
	v \in \mathcal{A} \\
	\bm{b} \in \mathcal{B}
	\end{subarray}}{\argmax}\left\lbrace\left|\bm{y}[k] - \mathbf{H} \bm{b} v \right|^2\right\rbrace
\end{align}
where $\mathcal{B}\subset \{ 1, 0\}^{N_{\rm t}}$ is the set of all vectors with a single entry $1$ and $N_{\rm t}-1$ entries $0$.

\subsection{Pulse Shaping for SM}
As pointed out in \cite{ishibashi2014effects}, switching antennas in each transmission interval distorts pulse shapes if the impulse response of the pulse shaping filter exceeds a symbol duration. To prevent excessive bandwidth occupation, the standard \ac{rrc} filter impulse response is truncated to some acceptable duration. The symbol rate is then reduced such that the pulses are separated in time in order to prevent intersymbol interference. \cite{ishibashi2014effects} suggests that an acceptable spectral shape is obtained by a rate reduction of factor $6$. The suggested approach deteriorates the effective spectral efficiency.

To design a pulse shape for \ac{sm}, we note that a square-root Nyquist pulse is not necessary if the symbols do not overlap in time-domain.
We hence propose using standard \ac{fir} low-pass filters, which are inherently time-limited.
Literature knows many distinct \ac{fir} filter design methods with different optimisation goals, such as linear-phase or minimum-order given a spectral mask with transition zones.
However, in a matched filter receiver the actual transfer function form of the pulse shaping filter is irrelevant as long as out-of-band emissions are suppressed to a certain level.
On the other hand, all energy which is transmitted out of band is inherently wasted even if a required minimum stopband attenuation is reached.
Hence, the non-Nyquist pulse shaping filter should maximise the ratio of energy inside the intended band (passband) and the total energy of its transfer function.
This criterion is fulfilled by the Slepian window which maximally concentrates the energy in the main lobe for a given window length $L$ \cite{6771595}.
The free design parameter $\alpha$ defines the width of the main lobe, whereas increasing the filter length reduces the side lobe level.

To compare the Slepian window with the truncated \ac{rrc} filter, we set the following requirements:
\begin{enumerate}
\item Pulse shaping is accomplished in digital domain.
\item The sampling rate is fixed for both pulse shaping filters to some $f_{\rm s}$. For the \ac{rrc} filter, filter parameter $T_0$ is set to\footnote{Note that $f_{\rm s} > 1/T_0$.} $4/f_{\rm s}$. Conventionally, this setting describes a system with symbol rate $1/T_0$ and fourfold oversampling for pulse shaping.
\item A common roll-off factor $\alpha = 0.4$ is set.
\item To limit emissions outside the bandwidth, a spectral mask is considered allowing a sidelobe level less than $-35$ dB.
\end{enumerate}

\begin{figure}[t]
\centering
\input{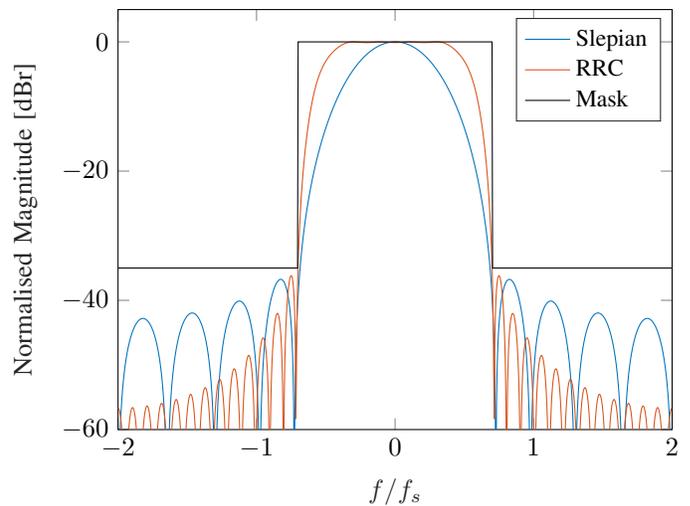}
\caption{Comparison of Slepian and RRC pulse shaping filters of length 10 and 37 samples, respectively. The spectral mask ensures a sidelobe level smaller than $-35$ dBr. }
\label{fig::pulseShapingSpectrum} 
\end{figure}

Fig.~\ref{fig::pulseShapingSpectrum} shows a Slepian window of length $10$ and $\alpha = 0.65$ as well as an \ac{rrc} filter of length $37$ samples. As the figure depicts, both filters fulfill the spectral mask. The spectral efficiency is reduced by $\zeta_{\rm S} = {2.5}$ and $\zeta_{\rm RRC}={9.25}$, if time-separated pulses of Slepian window and \ac{rrc} shape are transmitted instead of square-root Nyquist pulses. This indicates that the loss of spectral efficiency, introduced by the Slepian window, is almost four times smaller than the loss imposed by the truncated \ac{rrc} filter. We hence consider only Slepian pulse shaping in the remainder of this manuscript.

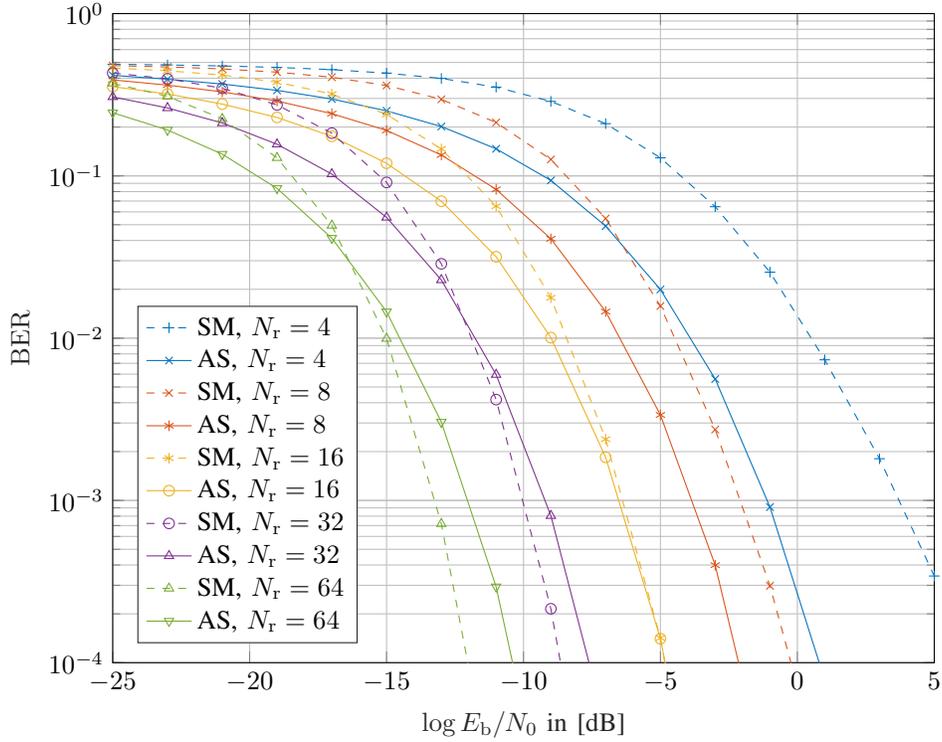
\begin{figure*}[!t]
\centering
%
%
\definecolor{mycolor1}{rgb}{0.00000,0.44700,0.74100}%
\definecolor{mycolor2}{rgb}{0.85000,0.32500,0.09800}%
\definecolor{mycolor3}{rgb}{0.92900,0.69400,0.12500}%
\definecolor{mycolor4}{rgb}{0.49400,0.18400,0.55600}%
\definecolor{mycolor5}{rgb}{0.46600,0.67400,0.18800}%
\begin{tikzpicture}

\begin{axis}[%
width= 4.3in,
height=3.4in,
at={(1.083in,0.573in)},
scale only axis,
xmin=-25,
xmax=5,
xlabel style={font=\color{white!15!black}},
xlabel={$\log \snr$ in [dB]},
ymode=log,
ymin=0.0001,
ymax=1,
yminorticks=true,
ylabel style={font=\color{white!15!black}},
ylabel={$\rm BER$},
axis background/.style={fill=white},
xmajorgrids,
ymajorgrids,
yminorgrids,
legend style={at={(0.03,0.03)}, anchor=south west, legend cell align=left, align=left, draw=white!15!black}
]
\addplot [color=mycolor1, dashed, mark=+, mark options={solid, mycolor1}]
  table[row sep=crcr]{%
-25	0.487357\\
-23	0.4826502\\
-21	0.47571425\\
-19	0.4662851\\
-17	0.45162495\\
-15	0.42979145\\
-13	0.3987906\\
-11	0.35222115\\
-9	0.2878599\\
-7	0.20982145\\
-5	0.12918285\\
-3	0.0645832\\
-1	0.02550525\\
1	0.0073531\\
3	0.0018076\\
5	0.0003422\\
};
\addlegendentry{SM, $N_{\rm r} = 4$}

\addplot [color=mycolor1, mark=x, mark options={solid, mycolor1}]
  table[row sep=crcr]{%
-25	0.416469125\\
-23	0.394312625\\
-21	0.3681645\\
-19	0.336494\\
-17	0.297693625\\
-15	0.251576125\\
-13	0.201380875\\
-11	0.146857125\\
-9	0.093713875\\
-7	0.049040875\\
-5	0.0199495\\
-3	0.005597875\\
-1	0.000909875\\
1	7.8125e-05\\
3	1.5e-06\\
5	0\\
};
\addlegendentry{AS, $N_{\rm r} = 4$}

\addplot [color=mycolor2, dashed, mark=x, mark options={solid, mycolor2}]
  table[row sep=crcr]{%
-25	0.47798535\\
-23	0.4691762\\
-21	0.4562001\\
-19	0.43577405\\
-17	0.40549825\\
-15	0.36063075\\
-13	0.29627025\\
-11	0.21326335\\
-9	0.1262386\\
-7	0.054618\\
-5	0.0157732\\
-3	0.00273355\\
-1	0.0002976\\
1	1.73e-05\\
3	1.4e-06\\
5	1.5e-07\\
};
\addlegendentry{SM, $N_{\rm r} = 8$}

\addplot [color=mycolor2, mark=asterisk, mark options={solid, mycolor2}]
  table[row sep=crcr]{%
-25	0.39035025\\
-23	0.362472875\\
-21	0.329541625\\
-19	0.289196625\\
-17	0.24181325\\
-15	0.190570125\\
-13	0.134517875\\
-11	0.08257225\\
-9	0.040947625\\
-7	0.01458675\\
-5	0.003363125\\
-3	0.000399\\
-1	1.5625e-05\\
1	1.25e-07\\
3	0\\
5	0\\
};
\addlegendentry{AS, $N_{\rm r} = 8$}

\addplot [color=mycolor3, dashed, mark=asterisk, mark options={solid, mycolor3}]
  table[row sep=crcr]{%
-25	0.46217335\\
-23	0.44450335\\
-21	0.4182706\\
-19	0.3782064\\
-17	0.31928955\\
-15	0.23921855\\
-13	0.14659865\\
-11	0.06502235\\
-9	0.0177362\\
-7	0.0023858\\
-5	0.000142\\
-3	1.65e-06\\
-1	0\\
1	0\\
3	0\\
5	0\\
};
\addlegendentry{SM, $N_{\rm r} = 16$}

\addplot [color=mycolor3, mark=o, mark options={solid, mycolor3}]
  table[row sep=crcr]{%
-25	0.354759375\\
-23	0.319781875\\
-21	0.276962\\
-19	0.22917975\\
-17	0.175341625\\
-15	0.119612375\\
-13	0.069822\\
-11	0.031727625\\
-9	0.010071375\\
-7	0.001843875\\
-5	0.0001405\\
-3	2.875e-06\\
-1	0\\
1	0\\
3	0\\
5	0\\
};
\addlegendentry{AS, $N_{\rm r} = 16$}

\addplot [color=mycolor4, dashed, mark=o, mark options={solid, mycolor4}]
  table[row sep=crcr]{%
-25	0.43092615\\
-23	0.3968732\\
-21	0.3466553\\
-19	0.2737428\\
-17	0.18332905\\
-15	0.09117\\
-13	0.0287211\\
-11	0.004182\\
-9	0.0002149\\
-7	2.75e-06\\
-5	0\\
-3	0\\
-1	0\\
1	0\\
3	0\\
5	0\\
};
\addlegendentry{SM, $N_{\rm r} = 32$}

\addplot [color=mycolor4, mark=triangle, mark options={solid, mycolor4}]
  table[row sep=crcr]{%
-25	0.306787125\\
-23	0.262210875\\
-21	0.21191575\\
-19	0.15669425\\
-17	0.1026455\\
-15	0.055462125\\
-13	0.02284075\\
-11	0.005960625\\
-9	0.00080475\\
-7	4.025e-05\\
-5	7.5e-07\\
-3	0\\
-1	0\\
1	0\\
3	0\\
5	0\\
};
\addlegendentry{AS, $N_{\rm r} = 32$}

\addplot [color=mycolor5, dashed, mark=triangle, mark options={solid, mycolor5}]
  table[row sep=crcr]{%
-25	0.37212075\\
-23	0.3093633\\
-21	0.2246113\\
-19	0.12930235\\
-17	0.0494708\\
-15	0.00994335\\
-13	0.00071795\\
-11	1.21e-05\\
-9	0\\
-7	0\\
-5	0\\
-3	0\\
-1	0\\
1	0\\
3	0\\
5	0\\
};
\addlegendentry{SM, $N_{\rm r} = 64$}

\addplot [color=mycolor5, mark=triangle, mark options={solid, rotate=180, mycolor5}]
  table[row sep=crcr]{%
-25	0.2449335\\
-23	0.191856\\
-21	0.136077\\
-19	0.083739125\\
-17	0.041347625\\
-15	0.014554125\\
-13	0.0030395\\
-11	0.00029375\\
-9	8.125e-06\\
-7	1.25e-07\\
-5	0\\
-3	0\\
-1	0\\
1	0\\
3	0\\
5	0\\
};
\addlegendentry{AS, $N_{\rm r} = 64$}

\end{axis}
\end{tikzpicture}%
\caption{Comparison of the power efficiency of \ac{as} and \ac{sm} with QPSK modulation and eight transmit antennas.\vspace*{5mm} }
\label{fig::asvsmQpsk} 
\end{figure*}
\section{Performance Comparison}
Using \ac{sm} for band-limited transmission, there exists a clear trade-off which indicates:
\begin{itemize}
\item On the one hand, \ac{sm} increases spectral efficiency by exploiting the information conveyed in the index of the selected antenna.
\item On the other hand, antenna switching in \ac{sm} causes~spectral regrowth which is mitigated by a drastic reduction~in the symbol rate. This is in contrast to \ac{as}, where antenna switching occurs only once per coherence interval and standard square-root Nyquist pulse shaping is used. 
\end{itemize}

Considering both approaches, a fair comparison of the performance requires the band-limitation constraint to be taken into account. In this respect, we investigate the performance of \ac{as} and \ac{sm} in terms of the average bit error rate when the same number of information bits are transmitted per coherence time. Our investigations show that for a fixed number of transmit antennas, \ac{sm} outperforms \ac{as} in low and moderate \ac{snr}s, when the number of receive antennas is relatively large. This observation suggests that in massive \ac{mimo} settings with reduced hardware complexity, \ac{sm} is an effective low-complexity scheme for uplink transmission. 

\subsection{Scenario 1: QPSK Transmission for AS and SM}
Fig.~\ref{fig::asvsmQpsk} depicts the average bit error rates, achieved under \ac{as} and \ac{sm}, in terms of power efficiency. For both approaches, the modulation scheme is set to \ac{qpsk}. This means that $x[k]\in\set{-1,+1, -\mathrm{i}, +\mathrm{i}}$. The power efficiency is quantified in terms of $\snr$ which is defined as the \ac{snr} divided by the number of bits transmitted in each channel~use.

Considering the Slepian window of length $10$, illustrated in Fig.~\ref{fig::pulseShapingSpectrum}, the symbol rate in \ac{sm} transmission is downscaled by factor $4/10 = 0.4$. To approximately compensate the reduced symbol rate, the number of transmit antennas is set for both approaches to
\begin{align}
N_{\rm t} = 2^{ \lfloor \Gamma_{\text{QPSK}}(\zeta_{\rm S}-1) \rceil } = 8,
\end{align}
where $\lfloor . \rceil$ denotes rounding to the nearest integer and $\Gamma_{\text{QPSK}} = 2$ bit per channel use is the spectral efficiency of \ac{qpsk}.
In this case, the total spectral efficiencies for \ac{as} and \ac{sm} considering a roll-off factor of $\alpha = 0.4$ and QPSK modulation read
\begin{subequations}
\begin{align}
	\Gamma_\mathrm{AS} &= \frac{\Gamma_{\text{QPSK}}}{1+\alpha} \approx 1.43\\
	\Gamma_\mathrm{SM} &= \frac{\lfloor \zeta_{\rm S}\Gamma_{\text{QPSK}}\rceil}{\zeta_{\rm S}(1+\alpha)} \approx 1.43 .
\end{align}
\end{subequations}
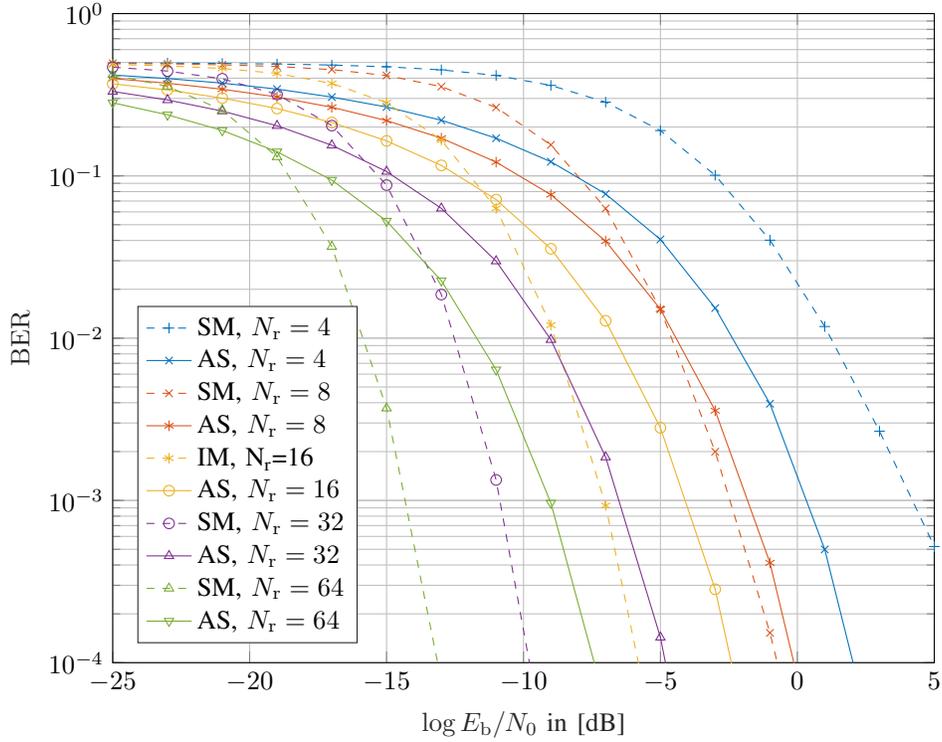
\begin{figure*}[!t]
\centering
%
%
\definecolor{mycolor1}{rgb}{0.00000,0.44700,0.74100}%
\definecolor{mycolor2}{rgb}{0.85000,0.32500,0.09800}%
\definecolor{mycolor3}{rgb}{0.92900,0.69400,0.12500}%
\definecolor{mycolor4}{rgb}{0.49400,0.18400,0.55600}%
\definecolor{mycolor5}{rgb}{0.46600,0.67400,0.18800}%
\begin{tikzpicture}

\begin{axis}[%
width=4.3in,
height=3.4in,
at={(1.083in,0.573in)},
scale only axis,
xmin=-25,
xmax=5,
xlabel style={font=\color{white!15!black}},
xlabel={$\log \snr$ in [dB]},
ymode=log,
ymin=0.0001,
ymax=1,
yminorticks=true,
ylabel style={font=\color{white!15!black}},
ylabel={$\rm BER$},
axis background/.style={fill=white},
xmajorgrids,
ymajorgrids,
yminorgrids,
legend style={at={(0.03,0.03)}, anchor=south west, legend cell align=left, align=left, draw=white!15!black}
]
\addplot [color=mycolor1, dashed, mark=+, mark options={solid, mycolor1}]
  table[row sep=crcr]{%
-25	0.4969015\\
-23	0.49525584375\\
-21	0.49310965625\\
-19	0.4888611875\\
-17	0.48203965625\\
-15	0.4703346875\\
-13	0.4500663125\\
-11	0.41606903125\\
-9	0.3611931875\\
-7	0.2847161875\\
-5	0.19021840625\\
-3	0.10087953125\\
-1	0.04014540625\\
1	0.01179325\\
3	0.0026715\\
5	0.00051990625\\
};
\addlegendentry{SM, $N_{\rm r}=4$}

\addplot [color=mycolor1, mark=x, mark options={solid, mycolor1}]
  table[row sep=crcr]{%
-25	0.41832775\\
-23	0.398124416666667\\
-21	0.372939583333333\\
-19	0.342704166666667\\
-17	0.3060345\\
-15	0.265722333333333\\
-13	0.220214333333333\\
-11	0.170614833333333\\
-9	0.122265833333333\\
-7	0.07738925\\
-5	0.040546\\
-3	0.0153226666666667\\
-1	0.00394225\\
1	0.000499083333333333\\
3	2.23333333333333e-05\\
5	1.66666666666667e-07\\
};
\addlegendentry{AS, $N_{\rm r}=4$}

\addplot [color=mycolor2, dashed, mark=x, mark options={solid, mycolor2}]
  table[row sep=crcr]{%
-25	0.49370075\\
-23	0.4898106875\\
-21	0.4836233125\\
-19	0.4722136875\\
-17	0.45179453125\\
-15	0.4156365\\
-13	0.3547303125\\
-11	0.26416921875\\
-9	0.155186375\\
-7	0.06299459375\\
-5	0.0150745625\\
-3	0.00199528125\\
-1	0.000152625\\
1	6.375e-06\\
3	9.375e-08\\
5	0\\
};
\addlegendentry{SM, $N_{\rm r}=8$}

\addplot [color=mycolor2, mark=asterisk, mark options={solid, mycolor2}]
  table[row sep=crcr]{%
-25	0.397883666666667\\
-23	0.3724855\\
-21	0.342084833333333\\
-19	0.306381583333333\\
-17	0.264842\\
-15	0.219092083333333\\
-13	0.170806416666667\\
-11	0.121581666666667\\
-9	0.0766189166666667\\
-7	0.03961475\\
-5	0.0150833333333333\\
-3	0.00357375\\
-1	0.0004125\\
1	1.6e-05\\
3	6.66666666666667e-07\\
5	0\\
};
\addlegendentry{AS, $N_{\rm r}=8$}

\addplot [color=mycolor3, dashed, mark=asterisk, mark options={solid, mycolor3}]
  table[row sep=crcr]{%
-25	0.4859614375\\
-23	0.47667821875\\
-21	0.4593431875\\
-19	0.42791815625\\
-17	0.37135071875\\
-15	0.282135\\
-13	0.1661095\\
-11	0.06306096875\\
-9	0.01202321875\\
-7	0.0009303125\\
-5	2.23125e-05\\
-3	0\\
-1	0\\
1	0\\
3	0\\
5	0\\
};
\addlegendentry{$\text{IM, N}_\text{r}\text{=16}$}

\addplot [color=mycolor3, mark=o, mark options={solid, mycolor3}]
  table[row sep=crcr]{%
-25	0.369540666666667\\
-23	0.338294916666667\\
-21	0.301634083333333\\
-19	0.260105333333333\\
-17	0.213384166666667\\
-15	0.164312166666667\\
-13	0.116020083333333\\
-11	0.0714145\\
-9	0.03554275\\
-7	0.01278525\\
-5	0.00280475\\
-3	0.000283416666666667\\
-1	8.25e-06\\
1	0\\
3	0\\
5	0\\
};
\addlegendentry{AS, $N_{\rm r}=16$}

\addplot [color=mycolor4, dashed, mark=o, mark options={solid, mycolor4}]
  table[row sep=crcr]{%
-25	0.467675125\\
-23	0.4426823125\\
-21	0.39650284375\\
-19	0.317752875\\
-17	0.20403465625\\
-15	0.08772228125\\
-13	0.018562875\\
-11	0.0013375\\
-9	1.88125e-05\\
-7	1.875e-07\\
-5	0\\
-3	0\\
-1	0\\
1	0\\
3	0\\
5	0\\
};
\addlegendentry{SM, $N_{\rm r}=32$}

\addplot [color=mycolor4, mark=triangle, mark options={solid, mycolor4}]
  table[row sep=crcr]{%
-25	0.33138025\\
-23	0.2937695\\
-21	0.250813333333333\\
-19	0.203651083333333\\
-17	0.154421666666667\\
-15	0.106371666666667\\
-13	0.0630786666666667\\
-11	0.0298363333333333\\
-9	0.00978316666666667\\
-7	0.001848\\
-5	0.00014325\\
-3	2.83333333333333e-06\\
-1	0\\
1	0\\
3	0\\
5	0\\
};
\addlegendentry{AS, $N_{\rm r}=32$}

\addplot [color=mycolor5, dashed, mark=triangle, mark options={solid, mycolor5}]
  table[row sep=crcr]{%
-25	0.42032096875\\
-23	0.35605384375\\
-21	0.25453759375\\
-19	0.130841125\\
-17	0.0366785625\\
-15	0.00369971875\\
-13	7.69375e-05\\
-11	0\\
-9	0\\
-7	0\\
-5	0\\
-3	0\\
-1	0\\
1	0\\
3	0\\
5	0\\
};
\addlegendentry{SM, $N_{\rm r}=64$}

\addplot [color=mycolor5, mark=triangle, mark options={solid, rotate=180, mycolor5}]
  table[row sep=crcr]{%
-25	0.282058666666667\\
-23	0.23784425\\
-21	0.19057\\
-19	0.140836583333333\\
-17	0.0941891666666667\\
-15	0.0526825\\
-13	0.0226088333333333\\
-11	0.00638483333333333\\
-9	0.00096425\\
-7	5.50833333333333e-05\\
-5	6.66666666666667e-07\\
-3	0\\
-1	0\\
1	0\\
3	0\\
5	0\\
};
\addlegendentry{AS, $N_{\rm r}=64$}

\end{axis}
\end{tikzpicture}%
\caption{Comparison of the power efficiency of \ac{as} with \ac{8qam} and \ac{sm} with \ac{qpsk} modulation and 64 transmit antennas.}
\label{fig::asvsm8qam} 
\end{figure*}

To investigate the impact of receive array size, the number of receive antennas is swept over $\mathcal{N}_{\rm r} = \set{8, 16, 32, 64}$. As Fig.\ref{fig::asvsmQpsk} shows, for all choices of $N_{\rm r}$, the \ac{as} approach initially performs superior to \ac{sm}. However, as the \ac{snr} grows, the curves for \ac{as} and \ac{sm} meet at some point from where \ac{sm} starts to outperform \ac{as}\footnote{For small values of $N_{\rm r}$, this crossover point occurs at high \ac{snr}s, and hence is not seen in the figure.}. As $N_{\rm r}$ grows, this crossover point moves to lower \ac{snr}s, e.g. for $N_{\rm r} = 32$, the point occurs around $\log \snr \approx -13$ dB.\vspace*{3mm}

The above observation is illustrated considering the following aspects:
\begin{itemize}
\item \textit{Aspect A:} The selection algorithm provides some diversity gain at the transmit side, and multiple receive antennas give diversity gain at the receive side. \ac{sm} however acquires diversity only at the receive side. This means that \ac{as} provides higher order of diversity compared to~\ac{sm}. 
\item \textit{Aspect B:} Using \ac{as}, signal recovery task is only symbol detection. However, \ac{ml} detection under \ac{sm} deals with both symbol and support recovery which can result in higher recovery error.
\item \textit{Aspect C:} With a fixed budget of energy for each channel use, the transmit symbol has higher energy under \ac{sm}, compared to the case with \ac{as}. In fact, in \ac{sm}, $\log_2 N_{\rm t}$ bits of information is encoded in the modulation index. These bits do not require energy allocation. Under \ac{as}, on the other hand, all information bits are transmitted by modulation over the active antenna, and hence the total energy is shared among all the transmitted bits.
\end{itemize}
At very low values of $\snr$, Aspects A and B dominate the performance, and hence \ac{as} outperform \ac{sm}. As $\snr$ increases, Aspect C starts to be the dominate aspect, and \ac{sm} becomes superior. Moreover, for a given $\snr$, the growth in the number of receive antennas makes the impacts of Aspects A and B insignificant. Hence, as $N_{\rm r}$ increases, \ac{sm} outperforms \ac{as} over a larger scope of \ac{snr}s. This implies that in massive \ac{mimo} setups, \ac{sm} is an efficient technique for uplink transmission.

\subsection{Scenario 2: 8QAM Transmission for AS and QPSK for SM}
The investigations on Scenario 1 is extended to the case with \ac{8qam} constellation for \ac{as} in Fig.~\ref{fig::asvsm8qam}.
For \ac{sm}, the conventional modulation, i.e.\ \ac{qpsk} remains unchanged.
Instead, the number of bits comprised by the indexing is increased to allow for optimum performance of \ac{sm}.
The spectral efficiency of \ac{8qam} is $\Gamma_{\text{8QAM}}=3$ bits per channel use. Hence, the loss in spectral efficiency, due to the reduction in symbol rate, weights more severe for \ac{sm}. Similar to Scenario 1, we compensate this loss by setting the number of transmit antennas
\begin{align}
N_{\rm t} = 2^{\lfloor \Gamma_{\text{8QAM}}\zeta_{\rm S}-\Gamma_{\text{QPSK}}\rceil} = 64 .
\end{align}
As a result, the total spectral efficiencies for the both approaches are given by
\begin{subequations}
\begin{align}
\Gamma_{\rm AS} &= \frac{\Gamma_{\text{8QAM}}}{1+\alpha} \approx {2.14}\\
\Gamma_{\rm SM} &= \frac{\lfloor \zeta_{\rm S}\Gamma_{\text{8QAM}}\rceil}{\zeta_{\rm S}(1+\alpha)} \approx {2.29}.
\end{align}
\end{subequations}

Fig.~\ref{fig::asvsm8qam} depicts similar behaviour as in Scenario 1. In this scenario, however, the crossover point occurs at lower \ac{snr}s compared to \ac{qpsk} transmission, the gap between the two approaches is larger. This follows from the fact that Aspect C, stated in previous scenario, becomes more significant as the size of transmit constellation set increases.
On the other hand, for a low number of receive antennas, Aspect B gets more dominant as a larger support has to be recovered by a the same small number of observations.

\subsection{Influence of Finite Switching Time}
In the above scenarios the switching time between the antennas was assumed to be negligible small.
This assumption may not hold for practical systems and switching time needs to be considered \cite{7420593}.
However, finite switching time influences \ac{as} and \ac{sm} and in different ways.
In \ac{as}, switching takes place only once during a coherence interval, whereas in \ac{sm} switching occurs with some probability after each symbol, depending on the data sequence \cite{7420593}.
In order to assure a deterministic (i.e.\ constant) symbol rate in \ac{sm}, the switching time $T_\mathrm{s}$ must be assumed to be a fixed part of the symbol duration. 
For completeness sake, it should also be mentioned that a non-negligible fraction of the coherence interval needs to be spent for channel estimation.
Since for \ac{as} and \ac{sm} full knowledge of \ac{csi} is assumed, we expect that this fraction is same in both cases and set it to zero for simplicity.
Then, the spectral efficiencies of \ac{as} and \ac{sm} for a common conventional modulation with spectral efficiency $\Gamma_{\text{mod}}$ are:
\begin{subequations}
	\begin{align}
	\Gamma_{\mathrm{AS}} &= \frac{\Gamma_{\text{mod}}}{1+\alpha}\left\lfloor \frac{T_\mathrm{c} - T_\mathrm{s}}{T_0} \right\rfloor  \frac{T_0}{T_\mathrm{c}}\\
	\Gamma_\mathrm{SM} &= \frac{\Gamma_{\text{mod}} + \log_2(N_{\rm t})}{1+\alpha} \left\lfloor \frac{T_\mathrm{c}}{\zeta_{\rm S} T_0 + T_\mathrm{s}}\right\rfloor \frac{T_0}{T_\mathrm{c}} .
	\end{align}
\end{subequations}
Note that in general $T_\mathrm{s}$ is much smaller than $T_\mathrm{c}$ such that $\Gamma_\mathrm{AS}$ is little influenced by the switching time.
In contrast, $\Gamma_\mathrm{SM}$ is essentially inversely proportional to $\zeta_{\rm S} T_0 + T_\mathrm{s}$. 
For example, if \ac{sm} shall achieve a comparable spectral efficiency as a conventional modulation of rate $25$ MSymb/s at some fixed $\Gamma_\text{mod}$ and $N_{\rm t}$, $T_\mathrm{s}$ needs to be significantly smaller than $\zeta_{\rm S} T_0 = 100$ ns.

\section{Conclusion}
It has been shown that a Slepian window is well suited as a pulse shaping filter for band-limited \ac{sm}, where standard square root Nyquist pulses cannot be used. As compared~to~a truncated \ac{rrc} filter, the shorter length of the Slepian window at a given maximum sidelobe level allows for a significantly higher symbol rate.

Using the Slepian window \ac{sm}, our comparison of \ac{as} and \ac{sm} under the constraint of equal band-limitation shows that \ac{sm} performs superior to \ac{as} for large receive arrays while \ac{as} outperforms \ac{sm} for receivers with a low number of antennas.
For high data rates, \ac{sm} requires a large number of transmit antennas to achieve comparable spectral efficiency as in \ac{as}, which eventually may lead to infeasible antenna array sizes. 
Contrary to \ac{as}, spectral efficiency of \ac{sm} is strongly affected by finite antenna switching time, so that the switching time ultimately sets an upper limit on the symbol rate in \ac{sm}. Therefore, \ac{sm} appears to be mostly suited for uplink scenarios in which moderate data rates are required.

\section{Acknowledgment}
The authors would like to thank Nikita Shanin for interesting discussions and his helpful comments on the work.

\bibliography{ref}
\bibliographystyle{IEEEtran}

\begin{acronym}
\acro{mimo}[MIMO]{multiple-input multiple-output}
\acro{csi}[CSI]{channel state information}
\acro{tdd}[TDD]{time division duplexing}
\acro{awgn}[AWGN]{Additive White Gaussian Noise}
\acro{iid}[i.i.d.]{independent and identically distributed}
\acro{sm}[SM]{spatial modulation}
\acro{bs}[BS]{Base Station}
\acro{as}[AS]{antenna selection}
\acro{lse}[LSE]{Least Squared Error}
\acro{rhs}[r.h.s.]{right hand side}
\acro{lhs}[l.h.s.]{left hand side}
\acro{wrt}[w.r.t.]{with respect to}
\acro{rs}[RS]{Replica Symmetry}
\acro{rsb}[RSB]{Replica Symmetry Breaking}
\acro{papr}[PAPR]{Peak-to-Average Power Ratio}
\acro{rzf}[RZF]{Regularized Zero Forcing}
\acro{snr}[SNR]{Signal-to-Noise Ratio}
\acro{rf}[RF]{radio frequency}
\acro{ml}[ML]{maximum likelihood}
\acro{rrc}[RRC]{root-raise-cosine}
\acro{bpsk}[BPSK]{binary phase shift keying}
\acro{qpsk}[QPSK]{quadrature phase shift keying}
\acro{8qam}[8QAM]{eight point quadrature amplitude modulation}
\acro{fir}[FIR]{finite impulse response}
\end{acronym}

\end{document}